\documentclass[aps,pra,twocolumn,letterpaper]{revtex4}

\usepackage{amsfonts, amssymb, amsmath, graphicx}

\DeclareFontFamily{OT1}{pzc}{}
\DeclareFontShape{OT1}{pzc}{m}{it}{<-> s * [1.10] pzcmi7t}{}
\DeclareMathAlphabet{\mathpzc}{OT1}{pzc}{m}{it}

\newcommand{\vect}[1]{\boldsymbol{\mathrm{#1}}}
\newcommand{\ket}[1]{\left|#1\right\rangle}
\newcommand{\bra}[1]{\left\langle#1\right|}
\newcommand{\expval}[1]{\left\langle#1\right\rangle}
\newcommand{\abs}[1]{\left\lvert#1\right\rvert}
\newcommand{\Eqref}[1]{Eq.~\eqref{#1}}
\newcommand{\Fig}[1]{Fig.~\ref{#1}}

\def\beq{\begin{equation}}
\def\eeq{\end{equation}}
\def\bes{\begin{equation*}}
\def\ees{\end{equation*}}
\def\bfig{\begin{figure}}
\def\efig{\end{figure}}

\def\ie{i.e., }
\def\ea{\emph{et al.}}
\def\eg{e.g., }

\def\prl#1#2#3{Phys.\ Rev.\ Lett.\ {\bf #1}, #2 (#3)}
\def\pra#1#2#3{Phys.\ Rev.\ A {\bf #1}, #2 (#3)}
\def\prb#1#2#3{Phys.\ Rev.\ B {\bf #1}, #2 (#3)}

\def\g2D{g_{\mathrm{2D}}}
\def\SP{\texttt{SP}}
\def\MIN{\texttt{MIN}}

\begin{document}
	\title{Fluctuations and correlations in rotating Bose-Einstein condensates}
	\author{Soheil Baharian}
	\author{Gordon Baym}
	\affiliation{Department of Physics, University of Illinois at Urbana-Champaign, 1110 W. Green St., Urbana, IL 61801}
	\date{\today}

	\begin{abstract}
		We investigate the effects of correlations on the properties of the ground state of the rotating harmonically-trapped Bose gas by adding Bogoliubov fluctuations to the mean-field ground state of an $N$-particle single-vortex system. We demonstrate that the fluctuation-induced correlations lower the energy compared to that of the mean-field ground state, that the vortex core is pushed slightly away from the center of the trap, and that an unstable mode with negative energy (for rotations slower than a critical frequency) emerges in the energy spectrum, thus, pointing to a better state for slow rotation. We construct mean-field ground states of 0-, 1-, and 2-vortex states as a function of rotation rate and determine the critical frequencies for transitions between these states, as well as the critical frequency for appearance of a metastable state with an off-center vortex and its image vortex in the evanescent tail of the cloud.
	\end{abstract}

	\maketitle

	\section{Introduction}
	A rotating ultracold harmonically-trapped Bose gas is predicted to pass through many exotic phases with increasing rotation rate (for a recent review, see Ref.~\cite{Cooper-ReviewArticle}). The mean-field description, which omits all correlations, predicts the zero-temperature ground state for a large number of particles to be a vortex lattice~\cite{ButtsRokhsar-Nature, MeanField}, and is in good agreement with current experiments~\cite{VortexLatticeInBEC}. However, exact diagonalization of the many-body Hamiltonian~\cite{Cooper-PRL-Melting} suggests the breakdown at very high rotation rates of the mean-field picture and a melting transition to strongly-correlated ground states, bosonic analogues of quantum Hall states~\cite{CorrelatedStates, Viefers-ReviewArticle}. The onset of correlations and quantum fluctuations can be expected to play a significant role in this transition to a strongly-correlated quantum liquid. However, a consistent theory of the zero-temperature melting of the vortex lattice does not exist so far~\cite{MeltingTheory-SinovaHannMacDonald, MeltingTheory-Baym, MeltingTheory-GhoshBaskaran, MeltingTheory-WuFengLi} (for a theory of thermal melting of the lattice, see Ref.~\cite{ThermalMelting}). A crucial first step in constructing such a theory is to understand better how correlations affect the system.

	With increasing rotation rate, the cloud expands in the transverse direction, and the particle density decreases. In each unit cell of the lattice, the vortex core occupies a larger fraction of the area of the cell~\cite{BaymPethick-PRA-VortexCore}, and the average displacement of the vortex from its equilibrium position increases~\cite{MeltingTheory-Baym} due to the zero-point motion of the Tkachenko mode~\cite{TkachenkoModes, TkachenkoModes-Baym}. Hence, the uncertainty in the position of vortices, which plays a leading role in the melting, increases at faster rotation rates~\cite{Cooper-ReviewArticle}. This uncertainty is completely absent in the mean-field picture, in which the vortex positions are fixed and do not fluctuate.

	The nature of the correlations between particles changes as the rotation rate increases. For angular momentum per particle less than or equal to unity (in units of $\hbar$ throughout), correlations in the exact ground state wave function~\cite{SmithWilkin-PRA-exactGS} are described by polynomials in the relative distance of the particles from the center-of-mass, $\psi\sim\sum_{i_1 < i_2 < \dots < i_L} (z_{i_1}-z_c)(z_{i_2}-z_c)\cdots(z_{i_L}-z_c)$ where $z \sim (x+iy)$ are the positions of the particles and $z_c$ the center-of-mass in the complex plane. On the other hand, when the angular momentum per particle is of order the number of particles, correlations appear in the distances of particles from each other, as in the bosonic Laughlin wave function~\cite{WilkinGunnSmith-PRL-interactions} $\psi\sim\prod_{i<j}(z_i-z_j)^2$.

	The aim of the first part of this paper is to build relevant correlations on top of the mean-field many-body ground state and to investigate their effects on the energetics and physical properties of the ground state. Based on the inferred modifications of the ground state, the second part of this paper investigates, at the mean-field level, different ground states of the ($0$-, $1$-, and $2$-vortex) Bose gas and their respective transitions as the external rotation rate increases.

	Small-amplitude Bogoliubov fluctuations around the mean-field ground state induce correlations by allowing small numbers of excitations to appear in nearby single-particle states. In a condensate with large number of vortices, the number of excited modes (single-particle harmonic oscillator eigenstates) involved is of order the number of vortices. Therefore, carrying out the general diagonalization is a mathematically challenging task for a many-vortex condensate. However, one can gain insight by working with a few-vortex system; for example, including the first three harmonic oscillator states is sufficient to describe systems with up to two vortices, as we show below. The simplest such system is a condensate with one singly-quantized vortex at the center of the trap, rotating at the critical frequency $\Omega_c$, at which the vortex becomes thermodynamically stable~\cite{ButtsRokhsar-Nature, LinnFetter-PRA}, and, hence, having unit angular momentum per particle.

	We find, indeed, that the correlations induced by Bogoliubov fluctuations lead to a better ground state in the thermodynamic limit, lower in energy than the mean-field one. We also see that the fluctuations drive the vortex core away from the center of the trap by a fluctuating distance of $\mathcal{O}(1/\sqrt{N})$ (in units of the characteristic length of the trap). Moreover, for rotations slower than the critical frequency $\Omega_c$, we find excitations with negative eigenenergy in the spectrum~\cite{LinnFetter-PRA} which remove one unit of angular momentum from the gas, indicating an instability towards a lower-energy non-rotating state for rotation rate $\Omega<\Omega_c$ and emphasizing the fact that the single-vortex mean-field ground state is not the best starting state.

	\bfig[t]
		\includegraphics[scale=0.425]{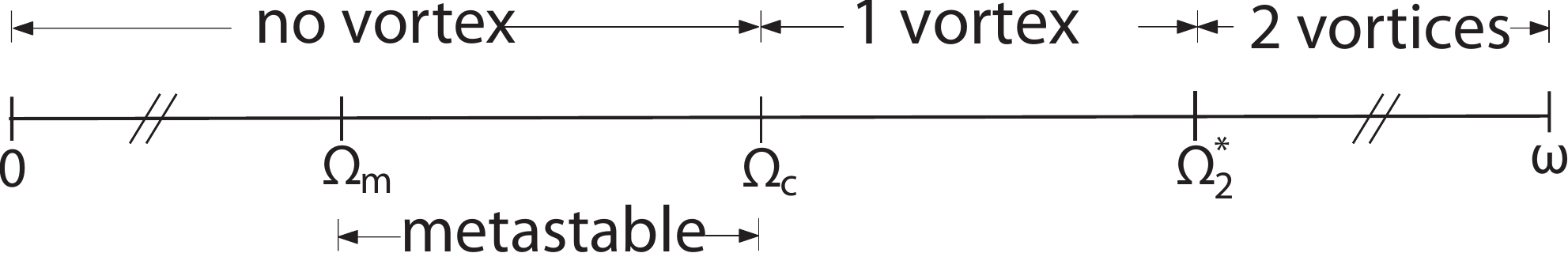}
		\caption{Schematic phase diagram of the condensate.\label{phasediagram}}
	\efig
	Based on these results, we construct a more energetically favorable mean-field condensate which, as a function of $\Omega$, is either non-rotating, a single-vortex state, or a two-vortex state. The phase diagram in \Fig{phasediagram} summarizes our results. At a certain frequency $\Omega_m$ (below $\Omega_c$), there exists a metastable state (a local minimum of the energy) with two off-center vortices in the cloud which are asymmetric about the origin; for a vortex close to the center of the trap, there exists an image vortex much further away where the particle density is negligible, in agreement with Ref.~\cite{NilsenBaymPethick-PNAS}. We calculate the critical frequencies, $\Omega_c$ and $\Omega^\ast_2$ respectively, at which the first and second vortices enter the cloud. The former agrees with the numerical result of Ref.~\cite{ButtsRokhsar-Nature}, while the latter is somewhat larger owing to our incorporating only a restricted number of single-particle eigenstates in the mean-field ground state.

	In the next section, we outline the basic description of the rotating Bose gas in terms of Landau levels in the Coriolis force. In Sec.~\ref{sec-Bogoliubov}, we determine the small-amplitude fluctuations about the mean-field condensate and their effect on the properties of the ground state, and then in Sec.~\ref{sec-StableCondensate}, we present the stable mean-field wave function that encapsulates the various phases of the rotating gas and their respective critical rotation frequencies. We set $\hbar = 1$ throughout.

	\section{Basic model}
	\label{sec-Theory}
	We consider a cloud of $N$ bosons of mass $m$ in a harmonic trap with frequencies $\omega$ in the $x$--$y$ plane and $\omega_z$ in the $z$ direction (with $\omega_z \gg \omega$ to tightly confine the gas in the axial direction), rotating around the $z$ axis with angular velocity $\Omega$. The characteristic oscillator lengths are $d= 1 / \sqrt{m\omega}$ in the transverse direction and $d_z = 1 / \sqrt{m\omega_z}$ in the axial direction. We assume weak two-body repulsive interactions of strength $g = 4 \pi a / m$ where $a$ is the s-wave scattering length. In the limit of fast rotation, $\Omega \lesssim \omega$, the gas becomes quasi-two-dimensional, and at zero temperature, it resides approximately in the ground state of the harmonic trap in the $z$ direction. The many-body Hamiltonian in the rotating frame is $\mathcal{H}^\prime = \sum^{N}_{i=1}(\mathpzc{h}^0_i - \Omega\mathpzc{l}_i) + \sum_{i<j}\g2D\delta(\vect{r}_i - \vect{r}_j)$, with the non-interacting single-particle Hamiltonian and angular momentum
	\beq
		\mathpzc{h}^0_i = \frac{\vect{p}^2_i}{2m} + \frac{1}{2}m\omega^2\vect{r}^2_i, \quad \mathpzc{l}_i = \hat{\vect{z}}\cdot(\vect{r}_i\times\vect{p}_i) ,
	\eeq
	where $\vect{r}$ and $\vect{p}$ are the particle positions and momenta in the $x$--$y$ plane, and $\g2D = g / \sqrt{2\pi}d_z$ is the effective interaction strength in two dimensions.

	The eigenstates of $\mathpzc{h}^0$ are the Landau levels $\ket{nm}$ -- where the Landau level index $n$ is the radial quantum number, and $m \geq -n$ is the angular momentum along the rotation axis -- with eigenvalues $\epsilon^\prime_{nm} = (2n+\abs{m}+1)\omega - m\Omega$. The characteristic energy scale of the two-body interaction is $V_0 = \g2D / 2\pi d^2$. The energy difference between two successive higher Landau levels ($n \neq 0$) is $\mathcal{O}(2\omega)$ which, for $\Omega\to\omega$, is much larger than that $\mathcal{O}(\omega-\Omega)$ between two states in a given Landau level. We assume the interactions to be sufficiently weak that $V_0 \ll 2\omega$; thus, we ignore the small higher Landau level components in the wave function and safely assume that the system resides in the manifold of the $n=0$ lowest Landau level (LLL) states. With this assumption, the only relevant quantum number is the angular momentum index $m$; from now on, we drop the Landau level index $n$ from the eigenfunctions, operators, and occupation numbers for simplicity, unless otherwise noted.

	The LLL eigenfunctions and eigenenergies are
	\begin{subequations}
		\begin{align}
			\phi_m(z) &= \langle\vect{r}\ket{0m} = \frac{1}{d\sqrt{\pi m!}}z^m e^{-\abs{z}^2 / 2}, \label{LLLeigenstate} \\
			\epsilon^\prime_{m} &= \epsilon^\prime_{0m} = \omega + (\omega-\Omega)m, \label{LLLeigenenergy}
		\end{align}
	\end{subequations}
	where $z=(x+iy)/d$ is the position in the complex plane; we use $\vect{r}$ and $z$ interchangeably. In terms of the corresponding creation and annihilation operators $a^\dagger_m$ and $a_m$, the second-quantized Hamiltonian in the rotating frame is
	\beq
		\label{H}
		\mathcal{H}^\prime = \sum_{m}\epsilon^\prime_{m}a^\dagger_{m}a_{m} + \frac{1}{2}\sum_{\substack{\lbrace m_i \rbrace}}\textrm{V}^{m_4 m_3}_{m_2 m_1} \, a^\dagger_{m_4}a^\dagger_{m_3}a_{m_2}a_{m_1}
	\eeq
	where the interaction matrix element in the LLL is
	\begin{align}
		\label{interactionMatrixElement}
		\textrm{V}^{m_4 m_3}_{m_2 m_1} &= \int \phi^\ast_{m_4}(\vect{r}) \phi^\ast_{m_3}(\vect{r}^\prime) \g2D\delta(\vect{r} - \vect{r}^\prime) \phi_{m_2}(\vect{r}^\prime)\phi_{m_1}(\vect{r}) \notag \\
			&= V_0 \, \frac{(m_1+m_2)! \, \delta_{m_3+m_4,m_1+m_2}}{2^{m_1+m_2}\sqrt{m_4!m_3!m_2!m_1!}}.
	\end{align}

	\section{Bogoliubov Hamiltonian in the lowest Landau level}
	\label{sec-Bogoliubov}
	We now turn to determining the effects of small-amplitude Bogoliubov fluctuations about the mean-field condensate on the properties of the system. We start with a condensate with a vortex at the center which rotates with angular frequency $\Omega_c$, to be determined. We derive an effective LLL Hamiltonian along with its excitation spectrum (which includes an unstable normal mode) and show that its ground state has lower energy than the initial mean-field state in which \emph{all} particles are condensed into the state $\ket{01}$. The initial condensate is
	\beq
		\psi(\vect{r}) = \sqrt{N_1} \, \phi_1(\vect{r})
	\eeq
	with $N_1$ particles in $\ket{01}$, describing a vortex at the center with winding number $1$. For $N_1 \lesssim N$, we make the usual replacement of the operators $a^\dagger_1$ and $a_1$, corresponding to $\ket{01}$, by $\sqrt{N_1}$ in the limit of large $N$. Although the total number of particles is fixed, interactions cause the number of particles in the condensate to fluctuate; the number of condensed particles can, thus, be written in terms of the total and the non-condensed particle numbers as
	\beq
		\label{N}
		N_1 = N - \sideset{}{'}\sum_m a^\dagger_m a_m
	\eeq
	where the prime indicates that $\ket{01}$ is excluded from the sum.

	In the thermodynamic limit ($N\to\infty$ with $NV_0$ constant), interaction terms that represent scattering of only one condensate particle or no condensate particles are respectively $\mathcal{O}(1/\sqrt{N})$ and $\mathcal{O}(1/N)$ smaller than those that involve two particles from the condensate and, thus, can be ignored. Following the standard procedure to write the Hamiltonian up to quadratic order in the excitation operators and using \Eqref{N} to conserve the total number of particles, we derive the LLL Hamiltonian in the rotating frame, 	
	\begin{align}
		\label{H-LLL-quadratic}
		\mathcal{H}^\prime = &\big[N(2\omega - \Omega) + \tfrac{1}{4}N^2V_0\big] + \big[\tfrac{1}{2}NV_0 - (\omega - \Omega)\big]a^\dagger_0 a_0 \notag \\
			&+ \big[\tfrac{1}{4}NV_0 + (\omega - \Omega)\big]a^\dagger_2 a_2 + \tfrac{1}{\sqrt{8}}NV_0\big(a^\dagger_0 a^\dagger_2 + a_0 a_2\big).
	\end{align}
	The constant first term in square brackets is the energy of the mean-field state with all $N$ particles condensed into $\ket{01}$ and no fluctuations present. The only two states in the LLL connected by the interactions in the presence of a condensate in $\ket{01}$ are $\ket{00}$ and $\ket{02}$, \ie for a LLL system, the maximum angular momentum transferred in any scattering process is $\pm 1$, whereas allowing higher Landau levels brings in and connects $\ket{10}$ and $\ket{12}$, $\ket{20}$ and $\ket{22}$, etc.  Larger transfers of angular momentum take the system out of the LLL as well, \eg $\ket{03}$ is connected to $\ket{1,-1}$ by a transfer of $\pm 2$ units, $\ket{2,-2}$ to $\ket{04}$ by a transfer of $\pm 3$ units, etc.

	Conservation of angular momentum is reflected in the fact that any scattering process involves simultaneous transfers of $+m$ and $-m$ units of angular momentum (relative to the condensate). The same method presented here was previously used by Linn and Fetter~\cite{LinnFetter-PRA} to include higher Landau levels perturbatively in the Bogoliubov excitation spectrum of this system. Also, Dodd \ea~\cite{Dodd-PRA} have used a similar argument to describe angular momentum conservation in this system in the presence of an external perturbation. Later, Rokhsar reinterpreted their argument to show~\cite{Rokhsar-PRL-CoreState} the existence of a negative energy excitation (the anomalous mode) with vortex core properties similar to those we find in Sec.~\ref{sec-Bogoliubov-subsec-Properties}.

	The canonical transformations to the bosonic quasiparticle operators
	\beq
		\label{quasiparticles}
		\begin{split}
			\alpha_{+1} &= u \, a_2 + v \, a^\dagger_0 \\
			\alpha_{-1} &= u \, a_0 + v \, a^\dagger_2
		\end{split}
	\eeq
	(with $u$ and $v$ real and positive) diagonalize the Hamiltonian, provided that $u^2 = 2$ and $v^2 = 1$. The new operators describe quasiparticles with $\pm 1$ units of angular momentum relative to the condensate. Thus, the LLL Hamiltonian in the rotating frame becomes
	\begin{align}
		\label{H-LLL-diag}
		\mathcal{H}^\prime = &\big[N(2\omega - \Omega) + \tfrac{1}{4}N^2V_0 - \tfrac{1}{4}NV_0\big] \notag \\
			&+ \left(\Omega-\Omega_c\right)\alpha^\dagger_{-1}\alpha_{-1} + \left(\omega-\Omega\right)\alpha^\dagger_{+1}\alpha_{+1},
	\end{align}
	where
	\beq
		\label{anomalousFrequency}
		\Omega_c = \omega-\frac{1}{4}NV_0.
	\eeq
	This equation shows that the Bogoliubov ground state (with no excited quasiparticles) has lower energy compared to the mean-field one by $-NV_0/4$. Also, the normal mode denoted by $-1$ has negative eigenenergy in the region $\Omega<\Omega_c$, indicating an instability in the system against being condensed into $\ket{01}$; this is the anomalous mode (see Ref.~\cite{FetterSvidzinsky-ReviewArticle} and references therein). Its existence shows that $\psi(\vect{r})$ is not the correct condensate for $\Omega<\Omega_c$ and, therefore, the Hamiltonian in \Eqref{H-LLL-diag} is not the correct one for this regime. As the rotation rate increases beyond $\Omega_c$, further LLL states beyond $\{\ket{00}, \ket{01}, \ket{02}\}$ come into play in the ground state, especially once two or more vortices enter the cloud (see Sec.~\ref{sec-StableCondensate}). Then, one must include Bogoliubov fluctuations around this new ground state in order to find the Bogoliubov Hamiltonian and its normal modes. For simplicity, we limit the discussion here to $\Omega=\Omega_c$ which corresponds to our starting point, a system fully-condensed in $\ket{01}$.

	In the manifold of the first three lowest Landau levels, the field operator for removing a particle at position $\vect{r}$ is
	\beq
		\label{fieldOperator}
		\Psi(\vect{r}) = \phi_0(\vect{r}) \, a_0 + \phi_1(\vect{r}) \, a_1 + \phi_2(\vect{r}) \, a_2.
	\eeq
	Inverting \Eqref{quasiparticles} gives
	\beq
		\label{particles}
		\begin{split}
			a_0 &= u \, \alpha_{-1} - v \, \alpha^\dagger_{+1} \\
			a_2 &= u \, \alpha_{+1} - v \, \alpha^\dagger_{-1},
		\end{split}
	\eeq
	and, thus, the expansion of $\Psi(\vect{r})$ in terms of the quasiparticles is
	\begin{align}
		\label{modeExpansion}
		\Psi(\vect{r}) = &\psi(\vect{r}) + \Big[u\,\phi_2(\vect{r})\alpha_{+1} - v\,\phi_0(\vect{r})\alpha^\dagger_{+1}\Big] \notag \\
			&+ \Big[u\,\phi_0(\vect{r})\alpha_{-1} - v\,\phi_2(\vect{r})\alpha^\dagger_{-1}\Big].
	\end{align}
	This is the mode expansion $\Psi(\vect{r}) = \psi(\vect{r}) + \sum_j\big[u_j(\vect{r})\alpha_j-v^\ast_j(\vect{r})\alpha^\dagger_j\big]$ in terms of the quasiparticles (see, \eg Ref.~\cite{Fetter-AnnalsOfPhysics}). Hence, the amplitudes for the $-1$ eigenmode are
	\beq
		\label{modeAmplitudes}
		\begin{split}
			u_{-1}(\vect{r}) &= u\,\phi_0(\vect{r}) \\
			v_{-1}(\vect{r}) &= v\,\phi^\ast_2(\vect{r}).
		\end{split}
	\eeq
	This mode, although having a negative eigenenergy for $\Omega<\Omega_c$, has a positive norm, since
	\beq
		\int\!\!\big[\abs{u_{-1}(\vect{r})}^2 - \abs{v_{-1}(\vect{r})}^2\big]d^2r = u^2-v^2 = 1,
	\eeq
	and is thus physical. Since ${\langle\vect{r}\ket{02}}^\ast = \langle\vect{r}\ket{2,-2}$, we arrive at the same amplitudes as derived up to zeroth order in $V_0$ in Sec. III of Ref.~\cite{LinnFetter-PRA}. We note, however, that the only two states that are mixed, in fact, are $\ket{00}$ and $\ket{02}$ which are in the LLL, and not $\ket{2,-2}$ which is a higher Landau level; up to the level of the approximation used in this article, the fluctuations reside solely in the LLL.

		\subsection{Properties of the Bogoliubov Ground State}
		\label{sec-Bogoliubov-subsec-Properties}
		We now investigate the stable ground state of the Hamiltonian, \Eqref{H-LLL-diag}, for $\Omega=\Omega_c$ and show that fluctuations drive the vortex away from the center of the trap and modify its velocity profile. The order parameter is $\psi(\vect{r})=\expval{\Psi(\vect{r})}$ or, in terms of annihilation operators, $\expval{a_1}=\sqrt{N_1}$ for the macroscopic condensate with $N_1$ particles in $\ket{01}$. The condensed state $\ket{N_1}$ is a coherent state that satisfies the eigenvalue equation $a_1\ket{N_1}=\sqrt{N_1}\ket{N_1}$ with the normalized solution
		\beq
			\ket{N_1} = e^{-N_1 / 2}\,e^{\sqrt{N_1}a^\dagger_1}\ket{\mathrm{vac}}
		\eeq
		where $\ket{\mathrm{vac}}$ is the vacuum. This state does not conserve particle number.

		The Bogoliubov ground state is determined by the condition that no quasiparticles be present, \ie
		\beq
			\label{BogoliubovGS-Definition}
			\alpha_{\pm 1}\ket{\textrm{G}}=0.
		\eeq
		Following the standard procedure (see, \eg Ref.~\cite{Huang-StatMech}), we find the normalized Bogoliubov ground state
		\beq
			\label{BogoliubovGS}
			\ket{\textrm{G}} = \frac{1}{\sqrt{2}} \, e^{-a^\dagger_2 a^\dagger_0 / \sqrt{2}}\ket{N_1}
		\eeq
		which has an expectation value of the angular momentum operator $\mathcal{L} = \sum_m m \, a^\dagger_m a_m$ given by $\bra{\textrm{G}}\mathcal{L}\ket{\textrm{G}}=N$.

		We now compare the lab-frame energy of $\ket{\textrm{G}}$ with that of the mean-field and exact ground states. The exact non-normalized many-body ground state for $2 \leq L \leq N$ is~\cite{SmithWilkin-PRA-exactGS} 
		\beq
			\label{exactGS}
			\psi^L_\textrm{x}(z_1 \dots z_N) \! = \sum_{i_1 < i_2 < \dots < i_L} \mspace{-20mu} (z_{i_1}-z_c)(z_{i_2}-z_c)\cdots(z_{i_L}-z_c)
		\eeq
		where $z_c=\sum^N_{i=1}z_i/N$ is the center-of-mass coordinate; we suppress the factor $\exp[-\sum^N_{k=1}\abs{z_k}^2 / 2]$ common to all $N$-particle LLL states from now on for brevity. This state has energy $E^L_\textrm{x} = (N+L)\omega + V_0N(N-1-L/2)/2$ in the lab frame. The mean-field ground state for $L=N$ with a vortex at origin,
		\beq
			\label{MFGS}
			\psi_\textrm{mf}(z_1 \dots z_N) = \prod^N_{i=1}\phi_{1}(z_i),
		\eeq
		has energy $E_\textrm{mf}=2N\omega+V_0N^2/4$ in the lab frame. Therefore, at $L=N$ (and, hence, at $\Omega=\Omega_c$), the Bogoliubov ground state lies exactly half-way in energy between the mean-field ground state and the exact one.

		A diagnostic of the structure of the vortex is the circulation around a closed contour $\mathcal{C}$ encircling the center,
		\beq
			\Gamma = \oint_\mathcal{C} \vect{v}(\vect{r}) \cdot d\vect{r},
		\eeq
		quantized in units of $2\pi\hbar/m$ for a quantum vortex. The velocity is $\vect{v}(\vect{r}) = \expval{\vect{j}(\vect{r})} / \expval{\rho(\vect{r})}$ where $\expval{\vect{j}(\vect{r})} = (\hbar/m) \textrm{Im} \expval{\Psi^\dagger(\vect{r})\vect{\nabla}\Psi(\vect{r})}$ is the expectation value of the current operator and $\expval{\rho(\vect{r})} = \expval{\Psi^\dagger(\vect{r})\Psi(\vect{r})}$ is that of the density operator. In the mean-field state, \Eqref{MFGS},
		\beq
			\vect{v}_\textrm{mf}(\vect{r}) = \frac{\hbar}{m} \, \frac{\hat{\vect{\theta}}}{r}
		\eeq
		which describes an irrotational superflow (except at the origin where the vortex is located) with circulation $\Gamma_\textrm{mf}=2\pi\hbar/m$. However, the Bogoliubov ground state, \eqref{BogoliubovGS}, gives
		\begin{align}
			\mspace{-12.5mu}
			\expval{\vect{j}(z)} \! &= \! \frac{\hbar}{m}\bigg[N\bigg(1-\frac{2}{N}+\frac{3}{N^2}\bigg)+\abs{z}^2\bigg] \abs{z} \, \frac{e^{-\abs{z}^2}}{\pi d^2} \, \hat{\vect{\theta}}, \\
			\mspace{-12.5mu}
			\expval{\rho(z)} \! &= \! \bigg[1 + N\bigg(1-\frac{2}{N}+\frac{3}{N^2}\bigg)\abs{z}^2 + \frac{\abs{z}^4}{2}\bigg] \frac{e^{-\abs{z}^2}}{\pi d^2}. \label{densityG}
		\end{align}
		Thus, in the limit of large $N$, the velocity field is
		\beq
			\label{velocity-r}
			\vect{v}_\textrm{G}(\vect{r}) = \frac{\hbar}{m} \, \frac{r\hat{\vect{\theta}}}{r^2 + \Delta(r)}
		\eeq
		where $\Delta(r) = (1-\frac{1}{2}r^4)/(N+r^2)$ is the correction due to quantum fluctuations. The circulation in $\ket{\textrm{G}}$ is then
		\beq
			\Gamma_\textrm{G}(\vect{r}) = \Gamma_\textrm{mf}\times\frac{r^4 + N r^2}{\frac{1}{2}r^4 + N r^2 + 1}.
		\eeq

		For large but finite $N$, we find $\Gamma_\textrm{G}/\Gamma_\textrm{mf} \sim 1$ (increasing from $1/2$ to $4/3$) for the large range $1/\sqrt{N} \lesssim r \lesssim \sqrt{N}$ while $\Gamma_\textrm{G}/\Gamma_\textrm{mf} \to 0$ as $r \to 0$ and $\Gamma_\textrm{G}/\Gamma_\textrm{mf} \to 2$ as $r\to\infty$. The vortex (at the center of the trap in mean-field) is now pushed off-center by quantum fluctuations, hence the vanishing circulation as the contour shrinks towards the origin. The off-center vortex fluctuates very close about the origin as shown by the circulation approaching its mean-field value as the contour radius expands past $\mathcal{O}(1/\sqrt{N})$. On the other hand, as the contour expands even further towards infinity, the circulation grows to twice its mean-field value, indicating the presence of an image vortex much further from the origin. These results agree with those of Sec.~\ref{sec-StableCondensate}. In the thermodynamic limit ($N\to\infty$), however, $\Gamma_\textrm{G}$ equals the quantum of circulation everywhere except at the origin (where it is zero) and at infinity (where it is twice the quantum of circulation); therefore, increasing number of particles suppresses quantum fluctuations of the vortex and leads to the vortex being driven less further from the center of the trap and the image vortex being driven more outward to infinity.

		\bfig[t]
			\includegraphics[scale=0.81]{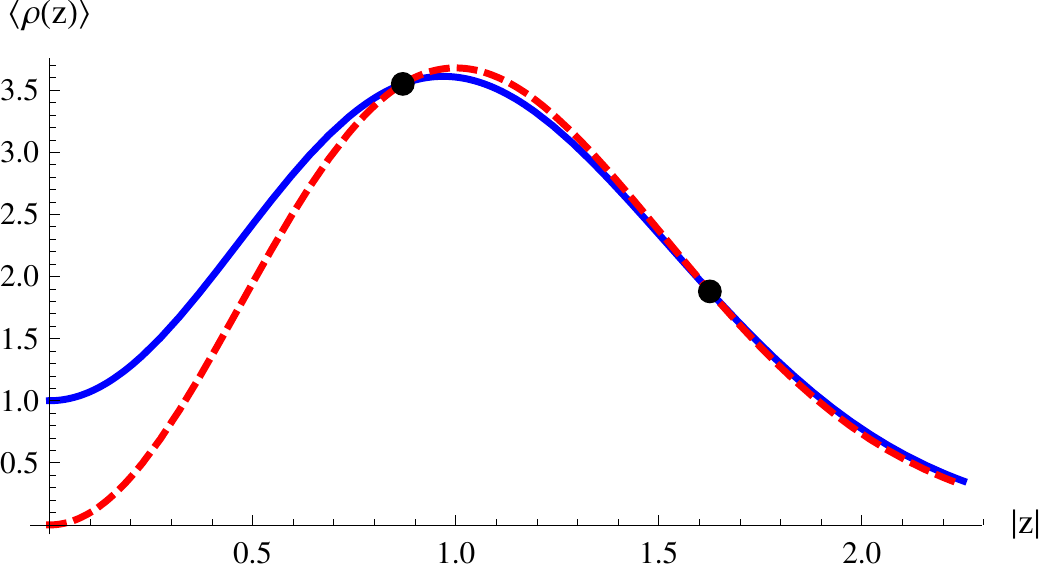}
			\caption{Density (in units of $1 / \pi d^2$) of the Bogoliubov (solid line) and mean-field (dashed line) ground states for $N=10$, showing a vortex at the center.\label{vortexcore}}
		\efig
		Quantum fluctuations change the particle density of the ground state compared to its mean-field value, $\rho_\textrm{mf} = N\abs{z}^2 e^{-\abs{z}^2} / \pi d^2$. The average density is now non-zero at the center of the trap, $\expval{\rho(0)}=1 / \pi d^2$ (see Fig. \ref{vortexcore}). One can understand this finite density in terms of the vortex fluctuating about the origin, as discussed above. Snapshots of the cloud in the laboratory would reveal a vortex at random positions (varying from shot to shot due to Bogoliubov fluctuations); averaging over these density snapshots would lead to \Eqref{densityG} for the average particle density of $\ket{\textrm{G}}$. One can also understand the finite density at the origin in terms of the single-particle quantum states, in particular, the non-zero occupation of $\ket{00}$ whose wave function does not vanish at the origin.

	\section{The Stable Condensate}
	\label{sec-StableCondensate}
	The anomalous mode, denoted by $-1$ in \Eqref{H-LLL-diag}, suggests that a condensate with a singly-quantized vortex at the center of the trap is not stable against fluctuations for $\Omega<\Omega_c$. However, Eqs. \eqref{modeExpansion} and \eqref{modeAmplitudes} indicate that a mean-field condensate wave function of the form
	\beq
		\label{stableCondensate}
		\psi(z)=\sqrt{N_1}\,\phi_1(z)+u\,\phi_0(z)-v\,\phi_2(z).
	\eeq
	can have lower energy and be stable, depending on the values of $u$ and $v$ (which can be taken to be real). We search for a better ground state by tuning the two extra degrees of freedom, $u$ and $v$, as follows.

	Normalizing $\psi(z)$ leads to $N=N_1+u^2+v^2$; hence, $u$ and $v$ are bounded by $u^2+v^2 \leq N$. The energy in the rotating frame, then, becomes
	\begin{align}
		\label{rotE}
		E^\prime = &\bigg[N(2\omega-\Omega)+\frac{1}{4}N^2V_0\bigg] + \bigg[\Big(\Omega-\omega+\frac{NV_0}{2}\Big)u^2 \notag \\
			&+ \Big(\omega-\Omega+\frac{NV_0}{4}\Big)v^2 - \frac{NV_0}{\sqrt{2}}uv\bigg] + V_0\Big(-\frac{1}{4}u^4 \notag \\
			&- \frac{5}{16}v^4 - \frac{3}{4}u^2v^2 + \frac{1}{\sqrt{2}}u^3v + \frac{1}{\sqrt{2}}uv^3\Big).
	\end{align}
	The constant term is the energy of a condensate with $N$ particles in $\ket{01}$. Note that $E^\prime$ is invariant under the simultaneous transformation $u \to -u$ and $v \to -v$. We choose $0 \leq v \leq u$ as this sector of the $u$--$v$ plane is energetically favorable.

	We denote $\tilde{E}^\prime = E^\prime - \big[N(2\omega-\Omega)+N^2V_0/4\big]$ as the energy contribution from the mixing of $\ket{00}$ and $\ket{02}$ with the condensate. Then, introducing the parametrization $u=\zeta\cosh(\theta/2)$ and $v=\zeta\sinh(\theta/2)$, we find
	\begin{align}
		\label{rotEexcitation}
		&\tilde{E}^\prime = \zeta^2\bigg[\Big(\Omega-\omega_\perp+\frac{NV_0}{8}\Big) \! + \! \frac{NV_0}{8}\big(3\cosh\theta-\sqrt{8}\sinh\theta\big)\bigg] \notag \\
			&\! + \! \zeta^4\frac{V_0}{128}\big[\! - \! 15 + 4\cosh\theta - 21\cosh(2\theta) + 16\sqrt{2}\sinh(2\theta)\big].
	\end{align}
	Ignoring the quartic part for now, \ie assuming $u,v \ll \sqrt{N_1}$ or $N_1 \lesssim N$, we minimize the quadratic part with respect to $\theta$ and find $\tanh\theta_m=\sqrt{8}/3$, which is depicted by the straight dashed line in the $u$--$v$ plane in \Fig{energyLandscape}. With this value of $\theta$, we have
	\beq
		\label{rotEexcitation-zeta}
		\tilde{E}^\prime = (\Omega-\Omega_c)\zeta^2 + \frac{3V_0}{16}\,\zeta^4.
	\eeq
	The quadratic term shows that up to second order in the mixing due to interactions, the system is unstable for $\Omega<\Omega_c$, as in the quantum treatment.

	For $\Omega>\Omega_c$, \Eqref{rotEexcitation-zeta} is a monotonically increasing function of $\zeta$ with a minimum at $\zeta=0$ or
	\beq
		u^>_m=v^>_m=0,
	\eeq
	describing a system fully condensed into $\ket{01}$ with one vortex at the center of the trap. (The superscripts {\lq\lq}$<${\rq\rq} or {\lq\lq}$>${\rq\rq} denote rotations slower or faster than $\Omega_c$.) The energy in the rotating frame becomes
	\beq
		E^{\prime >}_m = \Big(N\omega+\frac{1}{2}N^2V_0\Big) + N(\Omega_c-\Omega)
	\eeq
	where the first term is just the energy were all $N$ particles condensed into $\ket{00}$. For $\Omega\leq\Omega_c$, though, minimizing \Eqref{rotEexcitation-zeta} gives
	\beq
		\label{uvLess}
		u^<_m = \sqrt{\frac{16(\Omega_c-\Omega)}{3V_0}}, \quad\quad v^<_m = \sqrt{\frac{8(\Omega_c-\Omega)}{3V_0}},
	\eeq
	so that $u^<_m=\sqrt{2}\,v^<_m$, and the rotating-frame energy becomes
	\beq
		\label{rotEmLess}
		E^{\prime <}_m = \Big(N\omega+\frac{1}{2}N^2V_0\Big) + N(\Omega_c-\Omega) - \frac{4(\Omega_c-\Omega)^2}{3V_0}.
	\eeq

	The boundedness of $u$ and $v$ implies $0 \leq \zeta_m^2\cosh\theta_m \leq N$. Hence, the region of validity of Eqs. \eqref{uvLess} and \eqref{rotEmLess} is $\Omega_m\leq\Omega\leq\Omega_c$ where 
		\beq
		\label{approxOmegaMetastable}
		\Omega_m = \omega - \frac{3}{8}NV_0.
	\eeq
	Then, for $\Omega<\Omega_m $, the point $(u^<_m, v^<_m)$ lies outside the circle defined by $u^2+v^2=N$ and does not represent a physical solution.
	\bfig[t]
		\includegraphics[scale=0.75]{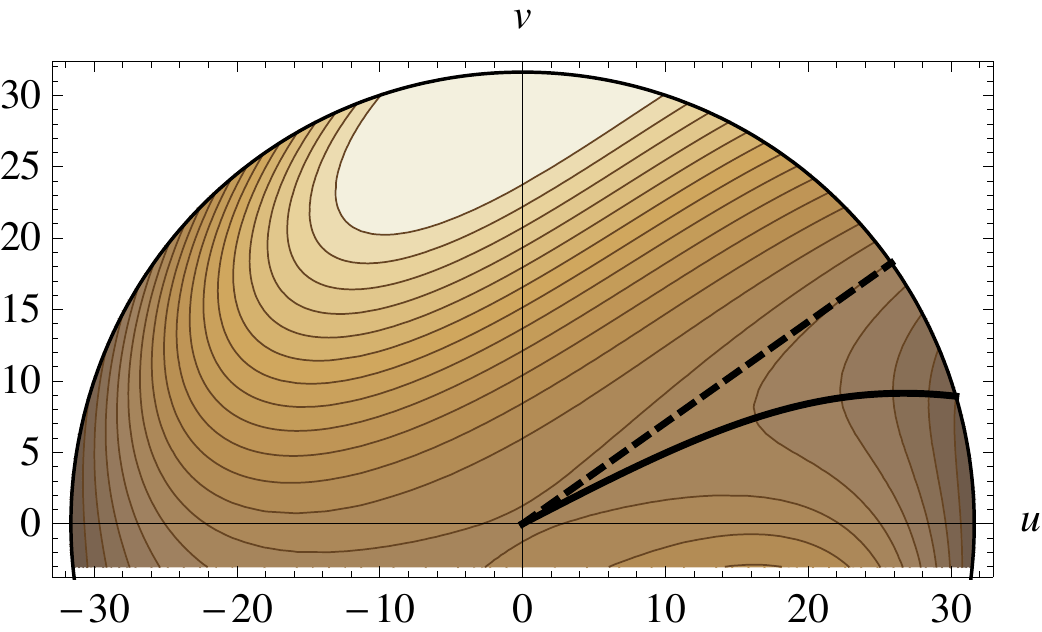}
		\caption{Contour plot of $E^\prime$ for $N=10^3$, $NV_0=0.1$, and $\Omega=\omega - \frac{3}{8}NV_0$. Darker shades indicate lower energies. The straight dashed line represents the direction given by $\tanh\theta_m=\sqrt{8}/3$ whereas the curved solid line is the solution of \Eqref{valleyEquation}, discussed below.\label{energyLandscape}}
	\efig

	In order to check for the existence of lower-energy states on the edge of the circle, where $u^2+v^2=N$, we have to compare $E^{\prime\lessgtr}_m$ with the corresponding energy $E^{\prime\lessgtr}_e$ for points on the edge. Since $N_1=0$ on the edge, we find
	\beq
		E^\prime_e = \Big(N\omega+\frac{1}{2}N^2V_0\Big) + 2(\Omega_c-\Omega)v^2 + \frac{3V_0}{16}v^4.
	\eeq
	For $\Omega\leq\Omega_c$, this energy increases monotonically with $v$, with the minimum at
	\beq
		u^<_e=\sqrt{N}, \quad\quad v^<_e=0,
	\eeq
	representing a system fully condensed into $\ket{00}$ (with no vortex) and the rotating-frame energy
	\beq
		E^{\prime <}_e = N\omega+\frac{1}{2}N^2V_0.
	\eeq
	For $\Omega>\Omega_c$, however, the minimum is at
	\beq
		u^>_e = \sqrt{N-\frac{16(\Omega-\Omega_c)}{3V_0}}, \mspace{27mu} v^>_e = \sqrt{\frac{16(\Omega-\Omega_c)}{3V_0}},
	\eeq
	and the energy in the rotating frame is
	\beq
		E^{\prime >}_e = \Big(N\omega+\frac{1}{2}N^2V_0\Big) - \frac{16(\Omega-\Omega_c)^2}{3V_0}.
	\eeq

	Therefore, for $\Omega_m\leq\Omega\leq\Omega_c$, the difference between the two energies is
	\beq
		E^{\prime <}_m - E^{\prime <}_e = \frac{4}{3V_0}(\Omega_c-\Omega)(\Omega-\Omega^\ast_1)
	\eeq
	where $\Omega^\ast_1 = \omega-NV_0$. Since $\Omega^\ast_1<\Omega_m$, we find that the point $(u=\sqrt{N},  v=0)$ corresponds to the \emph{global} minimum of the energy for the entire region $0\leq\Omega\leq\Omega_c$, indicating that the system has fully condensed into $\ket{00}$ with no vortex, whereas a \emph{local} minimum of the energy appears at $(u=u^<_m, v=v^<_m)$ for $\Omega_m\leq\Omega\leq\Omega_c$. This latter point corresponds to a metastable state which describes two vortices (asymmetric with respect to the origin) in the condensate; hence, the metastability frequency $\Omega_m$ is the rotation frequency at which a metastable state appears in the energy spectrum. On the other hand, for $\Omega>\Omega_c$, the difference between the two energies is
	\beq
		E^{\prime >}_m - E^{\prime >}_e = \frac{16}{3V_0}(\Omega-\Omega_c)(\Omega-\Omega^\ast_2)
	\eeq
	where $\Omega^\ast_2 = \omega-\frac{1}{16}NV_0$. Thus, if $\Omega_c<\Omega\leq\Omega^\ast_2$, the point $(u=0,v=0)$, \ie the center of the circle, corresponds to the \emph{global} minimum of the energy, indicating that the system has fully condensed into $\ket{01}$ with one vortex at the center. For $\Omega>\Omega^\ast_2$, however, the \emph{global} minimum of the energy is at $(u=u^>_e, v=v^>_e)$ on the edge of the circle, and the ground state is a coherent superposition of $\ket{00}$ and $\ket{02}$ with two vortices in a symmetric configuration with respect to the origin.

	Hence, $\Omega_c$ is the critical frequency for creating a centered vortex, and $\Omega^\ast_2$ is the critical frequency at which two vortices nucleate in the condensate. Note that $\Omega_c$ agrees with the external rotation frequency derived in Refs.~\cite{LinnFetter-PRA, ButtsRokhsar-Nature}; however, $\Omega^\ast_2$ is bigger than the two-vortex nucleation frequency, $\omega-0.078NV_0$, calculated numerically in Ref.~\cite{ButtsRokhsar-Nature} due to the very limited Hilbert space used here with only $\{\ket{00}, \ket{01}, \ket{02}\}$ as opposed to a rather large one used in Ref.~\cite{ButtsRokhsar-Nature}.

	The metastable state, for which $u^<_m=\sqrt{2}\,v^<_m$, has two vortices at the zeroes of $\psi(z)$, \ie
	\beq
		z^\pm_m(\Omega) = \frac{\sqrt{N-3(v^<_m)^2} \pm \sqrt{N+(v^<_m)^2}}{\sqrt{2}\,v^<_m}
	\eeq
	where $\Omega$ enters through $v^<_m$ on the right side. Using \Eqref{uvLess}, we find $v^<_m=\sqrt{N/3}$ at $\Omega=\Omega_m$. Hence, $z^\pm_m(\Omega_m)=\pm\sqrt{2}$. However, as $\Omega$ increases towards $\Omega_c$, we find for small $v^<_m$ that $z^\pm_m \to \pm\sqrt{2}(v^<_m / \sqrt{N})^{\mp 1}$ or, in other words, $z^+_m \to +\infty$ whereas $z^-_m \to 0$. As $\Omega$ increases from $\Omega_m$ to $\Omega_c$, the vortex at $+\sqrt{2}$ moves to infinity while the one at $-\sqrt{2}$ reaches the center and becomes stable there.

	It is instructive to compare this result to that of Ref.~\cite{NilsenBaymPethick-PNAS}, where the authors assume a condensate with an off-center vortex at position $b$ close to the center, \ie $\chi(z)\sim\sqrt{N}(z-b)$, and find perturbative corrections to the wave function using the Gross-Pitaevskii equation. The new non-normalized wave function in the LLL up to $\mathcal{O}(b^2)$ is $\chi(z) \sim \sqrt{N}(z-b)(1+bz/2+\cdots)$ (see Eq. (15) of Ref.~\cite{NilsenBaymPethick-PNAS}) with the rotation frequency, to lowest order in $b$, being $\Omega\simeq\Omega_c$ (see Eq. (16) of Ref.~\cite{NilsenBaymPethick-PNAS}). This wave function represents two vortices at $b$ and $-2/b$. Normalizing this wave function up to $\mathcal{O}(b^2)$ leads to
	\beq
		\chi(z) \sim \sqrt{N}\Big[\frac{b}{2}z^2 + \Big(1-\frac{3}{4}b^2\Big)z - b\Big].
	\eeq
	We can repeat a similar procedure for the metastable state for which $u^<_m=\sqrt{2}\,v^<_m$. Then, the condensate wave function, \Eqref{stableCondensate}, becomes
	\beq
		\psi(z) \sim \sqrt{N_1} \, z + u^<_m - \frac{u^<_m}{2} \, z^2.
	\eeq
	Expanding for small $u^<_m$ and defining $\mathpzc{b} = -u^<_m / \sqrt{N}$, we have
	\beq
		\psi(z) \sim \sqrt{N}\Big[\frac{\mathpzc{b}}{2}z^2 + \Big(1-\frac{3}{4}\mathpzc{b}^2\Big)z - \mathpzc{b}\Big].
	\eeq
	Thus, $\chi(z)$ and $\psi(z)$ have the same form with $b\leftrightarrow\mathpzc{b}$. Since $\mathpzc{b}$ is small, $\psi(z)$ also represents two off-center vortices at $\mathpzc{b}$ (close to the origin) and $-2/\mathpzc{b}$ (much further away in the evanescent tail of the cloud). Using \Eqref{uvLess}, we find the rotation rate of this two-vortex configuration in the lab frame to be $\Omega = \Omega_c - (3/16)NV_0\abs{\mathpzc{b}}^2$ which includes the next-order correction of $\mathcal{O}(\abs{\mathpzc{b}}^2)$ to the result of Ref.~\cite{NilsenBaymPethick-PNAS}. Note that while the calculations of Ref.~\cite{NilsenBaymPethick-PNAS} are limited to the fast rotating regime and are valid only in the vicinity of $\Omega_c$ (due to their perturbative nature in the small parameter $\Omega-\omega$), the method presented here covers the entire region $0\leq\Omega\leq\omega$ and is only limited by the number of states included in the condensate wave function.

		\subsection{The Valley and the Metastable Point}
		\begin{figure*}[t]
			\includegraphics[width=17cm]{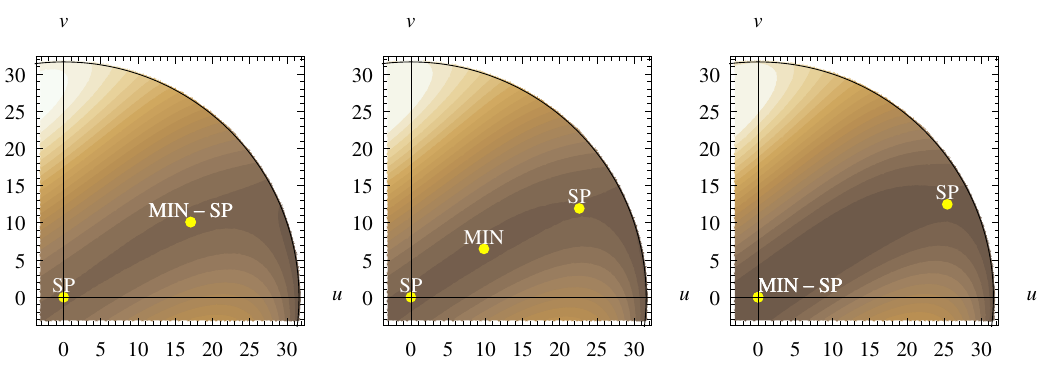}
			\caption{Energy landscapes for $\Omega = \omega - (2.1824/8)NV_0$, $\omega - (2.1094/8)NV_0$, and $\omega - (1/4)NV_0$ from left to right, for $N=10^3$ and $NV_0=0.1$. The dots represent the critical points with {\SP} and {\MIN} indicating the saddle-points and minima respectively. \label{criticalPoints}}
		\end{figure*}
		The derivation of the local minimum of the energy in the metastable regime, $\Omega_m\leq\Omega\leq\Omega_c$, has so far been restricted to small values of $u$ and $v$, \ie when $\Omega\to\Omega_c$ according to \Eqref{uvLess}. Ignoring the quartic term in \Eqref{rotEexcitation} leads to a constant $\theta_m$, representing a straight line in the $u$--$v$ plane which the local minimum traverses as $\Omega$ varies. Including the quartic terms causes the valley in the energy landscape to curve, as seen in \Fig{energyLandscape}. In this section, we rederive the metastable state and its onset frequency, $\Omega_m$, for larger values of $u$ and $v$, keeping the quartic terms in the energy.

		We first determine the equation governing the valley. The valley is a set of points, denoted here by $v(u)$, at which the change in the energy is extremum. We define $u=R\cos\eta$ and $v=R\sin\eta$ and write $\delta E^\prime = E^\prime(u+\delta u,v+\delta v)-E^\prime(u,v) \simeq (\delta u \, \partial_u E^\prime + \delta v \, \partial_v E^\prime)$. We find the direction that extremizes the change in the energy by keeping $R$ constant while varying $\eta$; thus, $\delta u = -R\sin\eta \, \delta\eta$ and $\delta v = R\cos\eta \, \delta\eta$. Then, $\delta E^\prime / \delta\eta = R(-\sin\eta \, \partial_u E^\prime + \cos\eta \, \partial_v E^\prime) = 0$ which gives $\partial_v E^\prime / \partial_u E^\prime = \tan\eta = \delta v / \delta u$, where the last equality is just the slope of the tangent to the curve $v(u)$. Therefore, the differential equation for the bottom of the valley is
		\beq
			\label{valleyEquation}
			\frac{dv}{du} = \frac{\partial_v E^\prime}{\partial_u E^\prime} \quad\quad \text{with} \quad\quad v(u=0)=0.
		\eeq
		Its solution for $\Omega=\omega-\frac{3}{8}NV_0$ is the solid line in \Fig{energyLandscape}.

		We rewrite $\tilde{E}^\prime$ as $\zeta^2 A(\theta)+\zeta^4 B(\theta)$ where
		\begin{align}
			A(\theta) \! &= \! \frac{NV_0}{8}\big[\tilde{\Omega}+(3\cosh\theta-\sqrt{8}\sinh\theta)\big] \label{A} \\
			B(\theta) \! &= \! \frac{V_0}{128}\big[\! - \! 15 \! + \! 4\cosh\theta \! - \! 21\cosh(2\theta) \! + \! 16\sqrt{2}\sinh(2\theta)\big] \label{B}
		\end{align}
		with $\tilde{\Omega} = (\Omega-\omega+\frac{1}{8}NV_0) / \frac{1}{8}NV_0$. At the critical points, $\partial_\zeta\tilde{E}^\prime=\partial_\theta\tilde{E}^\prime=0$. The trivial solution, $\zeta=0$, represents the center of the circle. Then, assuming $\zeta \neq 0$, we find
		\beq
			\label{criticalTheta}
			\frac{1}{A}\frac{\partial A}{\partial\theta} = \frac{1}{2B}\frac{\partial B}{\partial\theta}
		\eeq
		which gives the critical points for all values of $\Omega$.

		\begin{figure}[t]
			\includegraphics[height=8cm]{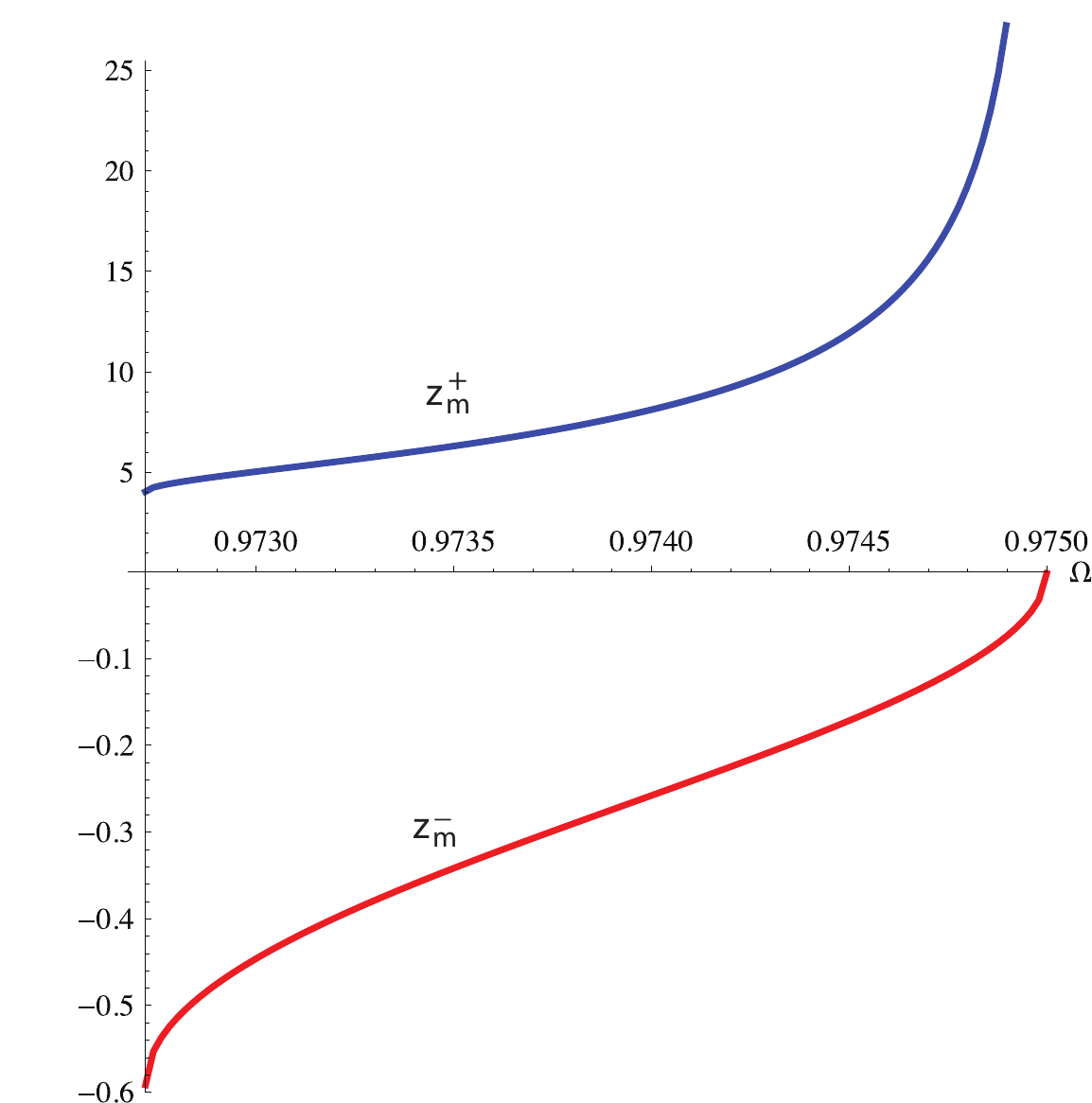}
			\caption{The position of the two off-center vortices for $N=10^3$, $NV_0=0.1$, and $\Omega_m\leq\Omega\leq\Omega_c$.  Note the different vertical scales for $z^+_m$ and $z^-_m$.\label{offCenterVortices}}
		\end{figure}

		For small $\Omega$, the only critical point in the valley is at the origin. However, \Fig{criticalPoints} shows that as $\Omega$ increases towards $\Omega_c$, two other critical points (a saddle-point and a minimum of the energy) appear in the valley. It is also evident that the minimum moves towards the center as $\Omega$ increases. Thus, there should be a rotation frequency, namely the metastability frequency $\Omega_m$, at which the saddle-point and the minimum lie on top of each other and, hence, the second derivative of the energy vanishes. Using \Eqref{criticalTheta}, we find that at $\Omega_m$,
		\beq
			\label{secondDerivative}
			\frac{1}{A}\frac{\partial^2 A}{\partial\theta^2} = \frac{1}{2B}\frac{\partial^2 B}{\partial\theta^2} - \left(\frac{1}{2B}\frac{\partial B}{\partial\theta}\right)^2.
		\eeq
		The single critical point (apart from the origin) at $\Omega_m$ satisfies both Eqs. \eqref{criticalTheta} and \eqref{secondDerivative}. Dividing \Eqref{secondDerivative} by \Eqref{criticalTheta} leads to an equation for $\theta$ independent of any other variable, namely
		\beq
			\frac{\partial^2 A / \partial\theta^2}{\partial A / \partial\theta} = \frac{\partial^2 B / \partial\theta^2}{\partial B / \partial\theta} - \frac{1}{2}\frac{\partial B / \partial\theta}{B},
		\eeq
		with the solution $\cosh\theta_m \simeq 2.0776$ at $\Omega_m$ or, using \Eqref{criticalTheta}, $\tilde{\Omega}_m \simeq -1.1824$. Therefore, the frequency at which the first metastable state appears is, in fact,
		\beq
			\label{exactOmegaMetastable}
			\Omega_m = \omega - \frac{2.1824}{8}NV_0
		\eeq
		which is much closer to the critical frequency $\Omega_c$ compared to the frequency given by \Eqref{approxOmegaMetastable}. Contour plots of energy for rotation frequencies $\Omega_m$ and $\Omega_c$ can be seen in the left and right panels of Fig.~\ref{criticalPoints}, showing that the saddle-point at the origin (for $\Omega<\Omega_c$) turns into a minimum for $\Omega>\Omega_c$. Since the approximations of the previous section are valid for $\Omega\lesssim\Omega_c$, the minimum approaches the origin following the line $v=u/\sqrt{2}$, \ie the slope of the valley at the origin is $1/\sqrt{2}$ near $\Omega_c$.

		The metastable state located at $(u_m, v_m)$ represents two off-center vortices at the zeroes of the condensate wave function, $z^\pm_m$, where
		\beq
			\frac{v_m}{\sqrt{2}}\,{z^\pm_m}^2 - \sqrt{N-(u^2_m+v^2_m)}\,z^\pm_m - u_m=0.
		\eeq
		The positions of the two roots of this equation as functions of $\Omega$ are plotted in \Fig{offCenterVortices} for the metastable regime, $\Omega_m\leq\Omega\leq\Omega_c$. Just as before, one vortex approaches the center of the trap while the other moves to infinity as $\Omega$ increases. However, their initial positions are not symmetric with respect to the origin but are at $z^-_m(\Omega_m) = -0.5917$ and $z^+_m(\Omega_m) = +4.2506$ for the particular values of $N$ and $V_0$ used in the figure. Since the vortices are stationary in the rotating frame, they precess around the origin with frequency $\Omega$ in the lab frame. Therefore, as seen in the lab frame, $z^-_m$ spirals in towards the center of the trap while $z^+_m$ spirals out to infinity as $\Omega$ increases.
		
	\begin{acknowledgements}
	This work was supported by the National Science Foundation under Grant Nos.~PHY07-01611 and PHY09-69790. We would like to thank C. J. Pethick for discussions leading to the general framework of the stable condensate. Also, S.B. would like to thank Mohammad Edalati and Esfandiar Alizadeh for helpful comments.
	\end{acknowledgements}

\end{document}